\documentclass[aps,pra,floatfix,showpacs,noshowkeys,epsfig,graphics,natbib]{revtex4-1}
\usepackage{graphicx}
\usepackage{amsmath}
\usepackage{amsfonts}
\usepackage{amssymb}

\newcommand{\newc}{\newcommand}
\newc{\beq}    {\begin{equation}}
\newc{\eeq}    {\end{equation}}

\newc{\beqa}    {\begin{eqnarray}}
\newc{\eeqa}    {\end{eqnarray}}
\newc{\bs}    {\section}
\newc{\no}    {\\ \nonumber}

\newc{\st}    {\stackrel}

\begin{document}
\title{ Gravity from Quantum Information }
\author{Jae-Weon Lee}\email{scikid@gmail.com}
%\affiliation{School of Computational Sciences,            Korea Institute for Advanced Study,             207-43 Cheongnyangni 2-dong, Dongdaemun-gu, Seoul 130-012, Korea}
\affiliation{ Department of energy resources development,
Jungwon
 University,  5 dongburi, Goesan-eup, Goesan-gun Chungbuk Korea
367-805}

\author{Hyeong-Chan Kim}
\email{hyeongchan@gmail.com}
\affiliation{School of Liberal Arts and Sciences, Korea National University of Transportation, Chungju, 380-702, Korea
}

\author{Jungjai Lee}
\email{jjlee@daejin.ac.kr}
\affiliation{Department of Physics, Daejin University, Pocheon, Gyeonggi 487-711, Korea}

\date{\today}

\begin{abstract}
It is suggested that the Einstein equation can be derived from %  by using
Landauer's principle applied to an information erasing process at a local Rindler horizon and Jacobson's idea linking the Einstein equation with thermodynamics.
When matter crosses the horizon, the information of the matter disappears and the horizon entanglement entropy increases to compensate the entropy reduction.
The Einstein equation describes an information-energy relation during this process, which implies that entropic gravity is related to the quantum entanglement of the vacuum and has a quantum information theoretic origin.
 \end{abstract}

\maketitle

\section{Introduction}

The Einstein equation express a relation between matter and the spacetime geometry  that the matter disturbs.
Despite of the clear geometric meaning of the Einstein tensor, the origin of this relation still remains a mystery.
In 1995, Jacobson suggested an interesting idea  that the
Einstein equation~\cite{Jacobson,Eling:2006aw} actually describes the equation of state (or,
the first law of thermodynamics) at local Rindler horizons.
Recently, Verlinde  brought us a remarkable new idea~\cite{Verlinde:2010hp} linking gravitational force to entropic force.
He derived the Newton's equation and the Einstein equation from the connection.
Padmanabhan also proposed that the Einstein equation is  from the equipartition law~\cite{Padmanabhan:2009kr}.

One can ask why there is such a surprising relationship between thermodynamics, especially entropy,  and gravity.
In this paper, we try to answer to this fundamental question by considering quantum information erasure in a curved spacetime.
%We present a related but slightly different idea connecting  gravity to information.
In a series of works~\cite{myDE,Kim:2007vx,Kim:2008re}
we  emphasized the  quantum informational nature of gravity.
For example,  we  suggested   that dark energy responsible for the
cosmic accelerating expansion is related to the quantum
entanglement of the vacuum fluctuation~\cite{myDE} or erasure
of  quantum information at a cosmic
horizon~\cite{forget,Lee:2008vn}.
A cosmic
horizon with a radius $R_h\sim O(H^{-1})$ has a kind of $thermal$
energy $E_h\propto T_h S_h \propto R_h$ associated with
its holographic entropy $S_h\propto R_h^2$, and this thermal
energy has an information theoretic origin. Here $H$ is the Hubble
parameter. (Recently, there appear similar suggestions based on
the Verlinde's idea~\cite{Li:2010cj,Zhang:2010hi,Wei:2010ww,Easson:2010av}.)
To be specific
we identified $S_h$ as an entanglement entropy $S_{Ent}$
associated with the erased vacuum information outside the horizon,
and $T_h\propto 1/R_h$ as the Hawking-Gibbons temperature of the
horizon. Then, it was straight forward to get a
horizon energy density $\rho_h\sim E_h/R_h^3\sim N_s
M_P^2/R_h^2$ which can be interpreted as  a holographic dark
energy density~\cite{li-2004-603}.
Here,  Planck mass $M_P$ is a UV-cutoff
and $N_s$  is the number of spin degree of freedom of quantum fields.
With reasonable input parameters
our dark energy model gives the energy density and the equation of state for dark energy comparable to  the observational data~\cite{myDE}.
(This dark energy could be also regarded as the energy of  cosmic Hawking radiation~\cite{Lee:2008vn}.)
%Let us start by briefly reviewing our previous works related to
%this subject.

Interestingly, in principle, one can explicitly
obtain  $S_{Ent}$ for the vacuum state
 using quantum field theory. %, not just a heuristic estimation.
 In ~\cite{Srednicki} $S_{Ent}$ for a spherical region is obtained by partial tracing
 a ground state of discretized quantum fields.
Using a similar approach we also
derived the first law of black hole thermodynamics from the second
law of thermodynamics~\cite{Kim:2007vx} and obtained a discrete black hole
mass formula. These works are based on  the Landauer's principle
in quantum information theory and the holographic principle.
 As a variant of the second law of thermodynamics,  Landauer's principle
states that to erase $N$ bits of information of a system irreversibly
at least $k_B N$ entropy of a bath should be increased and at least $k_B N T$
energy should be consumed, where $k_B$ is the Boltzman's constant
and $T$ is the temperature of the thermal bath contacting with the
system.
For a black hole and the universe their causal horizons play   roles of the
bath  as well as an information barrier.
We suggested that the horizons have thermal energy $E_h$ related to the information erasing at the horizons given by
\beq
\label{dE}
dE_h=k_B T_h dS_h,
\eeq
where $T_h$ is a horizon temperature and $d S_h$ is a horizon entropy
change due to the information erasing. We identified this energy as the origin of
 dark energy or black hole mass.
Thus, one can see that
this energy is very similar to the equipartition energy for the
entropic gravity~\cite{Padmanabhan:2009kr,Verlinde:2010hp}.
(Strictly speaking, the relation in Eq.(\ref{dE}) is more likely the Clausius relation, however
we call it the first law by convention.)

All our results above imply that there is an intrinsic  relationship between quantum information and gravity.
This can be seen as another realization of the famous slogan in quantum information community: ``It from Bit!"
Along this line in this paper we suggest that the Einstein equation itself can be
derived by considering  quantum information theory applied to Rindler horizons of a
given spacetime.
Our work is also based on the  work by Jacobson linking the first law of thermodynamics to the Einstein equation.
In Sec. II, we study the relation between entanglement and entropic gravity.
In Sec. III, we derive the Einstein equation from information loss.
In Sec. IV, we summarize and discuss the results.

\section{Entanglement and Entropic gravity}
In this section, we show that the entropic gravity can be related to the quantum entanglement and Landauer's principle applied to causal horizons.

The quantum entanglement is one of the key concepts of quantum information theory allowing
useful quantum information processes such as quantum key
distribution.
It  is a quantum nonlocal
correlation which can not be described by a classical correlation.
The entanglement entropy $S_{Ent}$ is a good measure of entanglement
for pure states such as the vacuum.
It is the von Neumann entropy $S_{Ent}=-Tr(\rho_A ln \rho_A)$
associated with the reduced density matrix $\rho_A\equiv Tr_B
\rho_{AB}$ of a bipartite system $AB$ described by a full density
matrix $\rho_{AB}$.
Why can we regard $S_{Ent}$ of the quantum field vacuum as $S_h$?
For
a causal horizon playing a role of information barrier, it is natural
to divide the system into
two subsystems A and B - inside and outside  the  horizon - and to
trace over one of the two regions to obtain the entanglement entropy of the horizon.
Thus, $S_{Ent}$ is ideal for $S_h$, when there is a causal horizon.

Using Landauer's principle, we calculated black hole mass
increase due to absorbtion of a test particle with energy $\delta
E$ in the context of quantum information theory in~\cite{Kim:2007vx}.
To the observer outside the black hole, this  corresponds to the erasure of $\delta S$
 bits of information by the thermal bath of the event horizon.
In this case the Landauer's principle demands that the entropy of the black hole increases at least
by $\delta S$ during this process.
Assuming this erasing process is optimal,  we obtained the increase of the black hole mass $
\delta M_{BH} = \delta E= k_BT\delta S,$
%\end{eqnarray}
which looks like the first law of black hole thermodynamics. Here, $T$ is the Hawking temperature for the black hole.
%This implies that the first law of black hole thermodynamics is
%just the second law disguised and information plays a crucial role in gravity.
%
In the work, to calculate the lost information
of the particle crossing the horizon
we  assumed an information erasing
 process similar to that  considered by Song and Winstanley~\cite{Song:2000fn}.
They derived a  generalized second law for black hole thermodynamics
from the view point of
quantum information theory, especially  by applying the Landauer¡¯s principle.
They considered a small
quantum system falling into the black hole with the Hawking temperature $T$. The system has Hamiltonian $H$,
density matrix $\rho_i$, and  energy $dE_{sys}=Tr(\rho_i H)$.
Around the horizon it
comes into thermal equilibrium with the black hole and its final state becomes  $\rho_f = e^{-H/k_B T}/Z$.
They showed that  the total change in entropy can be written as
%$\delta S_{sys}+\delta S_{BH}=tr[\rho_i ln \rho_i-\rho_f ln \rho_f]  \simeq   dE_{sys}/T\ge 0$.
\beq
\delta S\equiv \delta S_{sys}+\delta S_{BH}=tr[\rho_i \ln \rho_i-\rho_f \ln \rho_f] \simeq   dE_{sys}/k_BT \ge 0.
\eeq

Recently, a more accurate derivation of the entropy change is given by Padmanabhan (See chap 4.4 of ~\cite{Padmanabhan:2009vy}).
 Now, we apply his methods to calculate the variation of the entanglement entropy of the local Rindler horizon.
Consider a quantum density matrix $\rho_1$ of an excited state of a quantum field
crossing the Rindler horizon.
The density matrix can be obtained by tracing
out unobservable modes beyond the horizon.
This state has an entanglement entropy $S_1=-Tr(\rho_1 ln \rho_1)$.
Similarly defined vacuum state $\rho_0$ has an entanglement entropy $S_0=-Tr(\rho_0 ln \rho_0)$.
For the Rindler observer the vacuum state appears to be $\rho_0=e^{-H_R/k_B T_U}/Z$, where $H_R$ is
the corresponding  Hamiltonian in the Rindler coordinates and
 \beq
 T_U=\frac{\hbar a}{2\pi k_B c}
 \eeq
is the Unruh temperature of the horizon seen by an observer with acceleration $a$.
Following ~\cite{Padmanabhan:2009vy} one can find that the entanglement  entropy difference representing information loss
is
\beq
 \delta S_{Ent}\equiv S_1-S_0\simeq d E_{sys} /k_BT
 \eeq
  in the large acceleration limit.
Here, $dE_{sys}\equiv Tr((\rho_1-\rho_0) H_R)$.
Thus, we can generally use Eq. (\ref{dE}) to calculate the horizon entropy change
due to the particle crossing a causal horizon. Based on this result one can
 guess that the first law applied to causal horizons
 may have a quantum informational origin and the  information loss can be quantified in terms of the
  entanglement entropy $S_{Ent}$.

\begin{figure}[tpbh]
\includegraphics[width=0.4\textwidth]{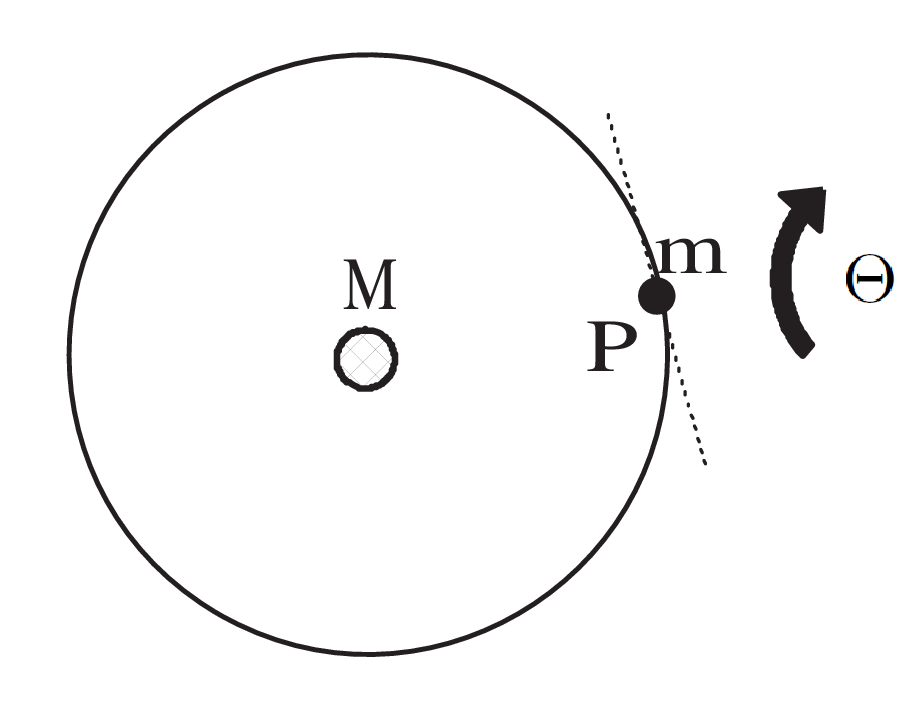}
\caption{To calculate the gravitational field at $P$  by a massive object
$M$ at the center,
consider an accelerating observer $\Theta$
 relative to the local inertial frame at $P$, where  a small test particle $m$ is.
 It is possible that for the observer
 the particle appears to  cross   a local Rindler horizon (represented by the dotted line) for the observer.
 This results in the increase of the entropy of the horizon and the Einstein equation describes the energy-information relation
 at this horizon.
}
\end{figure}

Putting it altogether, now it is natural to
imagine that gravity itself has a quantum informational origin. We
need to go one more step to confirm that. In 1995, Jacobson showed that one
can derive the Einstein equation  by demanding the first law of
thermodynamics,
\beq
\label{1}
 \delta Q = k_B T_U dS_h,
  \eeq
hold at Rindler horizons.
 Here, $S_h$ is the entropy of the horizon
and $\delta Q$ is the heat flux crossing the horizon.
He also assumed
the area law for the horizon entropy.
 %The bottom line of his theory is as follows.
By demanding that the first law holds at   local Rindler  horizons for
each spacetime point, one can derive the Einstein equation.
Below we follow his derivation keeping in mind   that
$dS_h= \delta S_{Ent} $ in Eq. (\ref{dE}) or Eq. (\ref{1}) actually represents the lost information when matter crosses the horizon.

 To understand this problem, consider  a massive object with mass $M$ at the center
 and an accelerating observer $\Theta$ with arbitrary large acceleration $a$
  relative to the local inertial frame at a point $P$ (See Fig. 1).
  There is another small test particle $m$ with energy $dE$ experiencing the gravitational field generated by $M$.
  The bottom line of our new idea is as follows.
When the test particle crosses the Rindler horizon, the information erasing
and apparent decrease of the total entropy happens to the observer.
%According to the Landauer's principle,
To save the second law of thermodynamics,
the horizon entropy should be increased at least by the amount of the information lost when the matter crosses the horizon.
We assume the optimal increase of the entropy.
The energy conservation gives $dE_h=dE$.
Hence, the amount of the entropy change is given by Eq. (\ref{dE}),
%\beq
$dS_h=\frac{dE_{h}}{k_B T_h}=\frac{dE}{k_B T_U}$.
%\eeq
This gives $dE=k_B T_U dS_h$.
From this relation at the Rindler horizon one can easily reproduce  Verlinde's formalism for entropic gravity (See ~\cite{mygravity} for details).
For example, using $dE\simeq mc^2$ and $T_U$ one can obtain
\beq
dS_h=\frac{dE}{k_B T_U}=\frac{2\pi c k_B m \Delta x}{\hbar},
\eeq
where $\Delta x\sim 1/a$ is the horizon distance.
This is just the entropy variation formula assumed by Verlinde.
It is also justified to invoke the Unruh temperature in his formalism.

%----------------------------------------------------
\section{Einstein equation from information loss   }
In this section, we show that the Einstein equation can be given from information loss and Jacobson's idea relating the gravity to thermodynamics.

Let us study more general situations in the context of  the quantum field theory in curved spacetime.
Thanks to the equivalence principle one can choose an approximately flat patch for every  spacetime point $P$
with boost Killing vector field $\xi_\alpha=-\kappa \lambda k_\alpha$ generating a
 local Rindler horizon for the observer $\Theta$ where the matter crosses. Here,
$k_\alpha$ is the tangent vector to the horizon generators with an affine parameter $\lambda$
and $\kappa$ is the acceleration of the Killing orbit satisfying $T_U={\hbar\kappa }/{2\pi k_B c}$.
Then, the local situation in Fig.1 at $P$
 can be approximately described using the Rindler coordinate chart in
Fig. 2;
\beqa
ct=\left( \frac{c^2}{a}+X \right) \sinh\left( \frac{a \tau}{c} \right), \no
x=\left( \frac{c^2}{a}+X \right) \cosh\left( \frac{a \tau}{c} \right)
\eeqa
where $(t,x)$ is the Minkowski coordinate
and $(\tau,X)$ is the Rindler coordinate of the observer.
Each point in the figure represents a 2-dimensional spatial plane.

\begin{figure}[tpbh]
\includegraphics[width=0.3\textwidth]{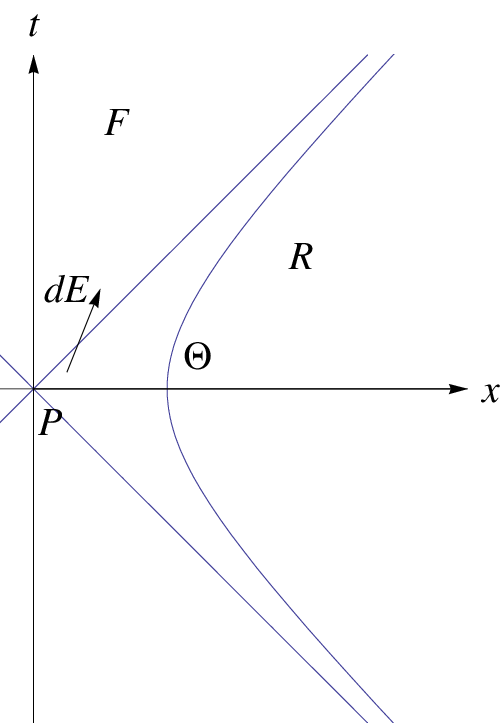}
\caption{To calculate the metric evolution at $P$ consider the
approximate Rindler chart for  the observer $\Theta$,
who would see the matter with energy $dE$ crosses the local Rindler horizon.
The horizon divides the spacetime into two causally disconnected  regions $F$ and $R$,
and should expand appropriately to satisfy the Landauer's principle.
This leads to the Einstein equation
}
\end{figure}

We can now follow Jacobson's derivation
and  generalize the first law  by defining the
energy flow across the horizon $\Sigma$
\beq
\label{generalized}
dE=-\kappa \lambda\int_\Sigma T_{\alpha\beta} \xi^\alpha d \Sigma^\beta
%=-\lambda \kappa\int_H T_{\alpha\beta} \xi^\alpha  \xi^\beta d\lambda dA
\eeq
where $d\Sigma^\beta=\xi^\beta d\lambda dA$, $dA$ is the spatial area element,
and $T_{\alpha\beta}$ is the energy momentum tensor of  the matter field.
Using the Raychaudhuri equation one can
denote the horizon area expansion $\delta A\propto dS_h$  and the increase of the entropy
as
\beq
\label{dA}
dS_h=\eta \delta A=-\eta \lambda \int_\Sigma R_{\alpha\beta} \xi^\alpha  d\Sigma^\beta,
\eeq
with some constant $\eta$ ~\cite{Jacobson}.
If $S_h$ saturates the Bekenstein bound,  $\eta=c^3/4\hbar G$.

It is well known that
by partial tracing the Minkwoski vacuum of matter field $\phi$, one can obtain a reduced density matrix $\rho_0=e^{-H_R/k_B T_U}/Z$ for the Rindler vacuum, which is a thermal one. (See for example \cite{susskindbook}.)
Here, $Z$ is the partition function.
This can be done by calculating the wave functional of fields in the left and right
 patches $(\phi_L,\phi_R)$ respectively with the path integral
\beq
\psi(\phi_L,\phi_R)\propto\int d\phi e^{-S_E},
\eeq
where $S_E$ is the Euclidean action.
Then, a reduced density matrix is given by $\rho_R=\exp(-2\pi H_R)/Z$, from which one can obtain $\rho_0$.

At this moment, if we identify the horizon entropy $S_h$ to be the entanglement entropy
$S_{Ent}$,
we can avoid the use of the holographic principle
and some circular logics. Furthermore, in principle one can calculate $\eta$ using
the quantum field theory  by
 adding up contributions from all fields~\cite{PhysRevD.52.4512}.
 Therefore, we conjecture that the horizon entropy variation connected to Landauer's principle
 is just $dS_{Ent}$.
 Although the initial state is not exactly a vacuum state, after the matter crosses the horizon, we can use the entanglement entropy of the vacuum
to describe the whole system. This is because, after crossing the horizon, the matter becomes a part
of the thermal system (i.e., the vacuum for an inertial observer) beyond the horizon to the outside Rindler observer.

  For example, for a spherical horizon with a radius $r$,
 if there are $N_j$  spin degrees of
freedom  of  the $j$-th field,
this implies that for a spherical region with radius $r$
\beq
S_{Ent}=\sum_j \beta_j N_j\frac{ r^2}{L_P^2}={4\pi \eta   r^2}.
\eeq
where $\beta_j$ is an $O(1)$ numerical constant for the $j$-th field and  $L_P= \sqrt{8\pi \hbar G/c^3}$
is the reduced Planck length.
Srednicki obtained  a value $\beta_j=0.3$ for
the massless  scalar field by performing numerical calculations on a sphere lattice.
A similar value was obtained for a  massless scalar field in the
Friedmann universe  in  ~\cite{PhysRevD.52.4512,PhysRevLett.71.666}.
Thus, the Bekenstein bound gives a constrain  $\Sigma_j \beta_j N_j= 8\pi^2$
on the number and characteristics
of quantum fields in the universe~\cite{myDE}.

To explain all horizon entropy  with $S_{Ent}$, $N_j$ should have a special value. However,
despite of this species problem of entanglement entropy, we asserts here that the entanglement entropy
is the origin of the horizon entropy needed for thermodynamic gravity models.
From the information theoretic viewpoint described above,
the entanglement entropy of the vacuum is the most natural candidate for $S_h$ related to Landauer's principle.
(After completion of the first version of this work, using perturbative quantum gravity, Bianchi indeed showed that the change of entanglement entropy due to the small matter
is equivalent to the change of
 the Bekenstein-Hawking entropy
~\cite{Bianchi:2012br}. This result
strongly supports our conjecture.)

Inserting Eqs. (\ref{generalized}) and (\ref{dA}) into $ dE =k_B T_U dS_{Ent}=\hbar \kappa dS_{Ent}/2\pi c$ one can see
$2\pi c T_{\alpha\beta} \xi^\alpha d \Sigma^\beta
=\eta  R_{\alpha\beta} \xi^\alpha d \Sigma^\beta$. For all local Rindler horizons this equation
should hold. Then, this condition and Bianchi identity lead to
the Einstein equation
\beq
\label{einstein}
R_{\alpha\beta}-\frac{R g_{\alpha\beta}}{2}+\Lambda g_{\alpha\beta}
=\frac{2\pi }{\eta c} T_{\alpha\beta }
\eeq
with the cosmological constant $\Lambda$ as shown in his paper.
Results in ~\cite{Bianchi:2012br} implies that the conjecture $dS_h=dS_{Ent}$ gives the correct value for $\eta$.

%-----------------------------------------------------------
\section{Discussions}
Strangely, information theoretic aspects of Jacobson's
 theory were not discussed widely so far, though the entropy $S_h$  clearly has  an information theoretic meaning
in modern physics (See ~\cite{Chirco:2009dc}).
Now, we can interpret  the first law in Eq. (\ref{1})
 in terms of the Landauer's principle. We assumed
 the second law of thermodynamics
is more fundamental than the first one and
the first law should be satisfied if the second law holds for all causal horizons blocking information.

There is yet another support for quantum information being the source of entropic gravity.
Using Levy's Lemma it was shown that if the universe is sufficiently large and  in a generic  pure  quantum
state, a small system entangled with environment in the universe can be effectively thermalized
~\cite{popescu}.

Our theory reproduces Verlinde's entropic gravity in terms of quantum information.
 How can one reconcile the irreversibility of the entropy and reversibility of gravity in our interpretation?
As in black hole cases, free falling observer comoving with free falling matter would not see the Rindler horizons, while
fixed  observers (accelerating against the matter frame relatively) can see the horizons.
Thus, the entropy %irreversibility
in a gravitational system is an observer dependent quantity in general.
Nonetheless, since the Einstein equation derived above
is covariant, the equation should hold for every frame, once it is satisfied in a specific frame.
Therefore, our theory is in concordance with Verlinde's proposal in a general sense.

However, there are some differences between our theory and Verlinde's theory.
First, in our theory we  assumed neither the proportionality of entropy on the distance,
nor the entropic force. The equipartition condition is not necessary either.
Second, we suggested the horizon entropy is originated from
 quantum information erasing at a horizon rather than coarse graining of mysterious microscopic degree of freedom on the horizon.
 This explain why the derivation of classical gravity is involved with $\hbar$
 and why gravity has something to do with entropy or information.
The Newton's gravity  could arise, of course, from the non-relativistic limit of the Einstein equation.
Third, our theory does not demand the generalized holographic principle for equipotential surfaces.
Ordinary quantum field theory in curved spacetime is enough to calculate the entanglement entropy of the horizons.
In principle, with reasonable assumptions,
 one may explicitly calculate some relevant physical quantities such as dark energy.
Note that we did not assume that the spacetime is emergent but
assume the existence of  spacetime a priori.
Since our theory links quantum mechanics to classical gravity,
it might  provide us a new way to quantum gravity.

Considering the second law, we expect that the causal horizon area of the universe
has a strong tendency to increase~\cite{Kim:2008re}. That is, matter in the universe distribute themselves
so that the horizon entropy of the universe to be maximized.
This might be the origin of the gravitational force as an entropic force considered by Verlinde.

Summarizing all these facts, we can say that
the Einstein equation simply states that
total entropy of matter and horizon should not decrease
and this is the bottom line of all gravitational phenomena.
The causal structure of the spacetime should be automatically arranged so that
the area of the Rindler horizons
  appropriately increase to compensate the information loss
of the matter crossing the horizons.

In short, the Einstein equation links matter to gravity
and his famous formula $E=mc^2$ links matter to energy.
We know also that the Landauer's principle links  information to  energy.
Thus, now we may have a relation between information and gravity, the Einstein equation with the
quantum information theoretic interpretation.
%Our theory implies that the Einstein equation is more about information
%rather than energy or equation of state.
%In other words, information might be more profound physical entity than matter or field.

\section*{acknowledgments}
Authors are thankful to  Gungwon Kang for
helpful discussions.
This work was supported in part by Basic Science Research Program through the
National Research Foundation of Korea (NRF) funded by the ministry of Education, Science and Technology
(2010-0024761), by the Topical Research Program (2010-T-01)
of Asia Pacific Center for
Theoretical Physics, and by the Korea Research Foundation Grant funded by the Korean
Government (MOEHRD, Basic Research Promotion Fund)(KRF-2008-314-C00063),
and by the Korea Research Foundation (KRF) grants funded by the Korea government (MEST) No. 2010-0011308 (H.K.)

%\bibliographystyle{prsty}
%\bibliography{entanglement}

\end{document}